# ICANN, "Internet Stability," and New Top Level Domains

Jonathan Weinberg[*]

Since 1998, an entity known as the Internet Corporation for Assigned Names and Numbers (ICANN) has administered the Internet domain name system. In November 2000, the ICANN board of directors agreed to add seven new top level domains to the name space. ICANN staff then embarked upon extensive negotiations with representatives of the registries that would operate the new domains, with the goal of signing agreements describing nearly every aspect of the registries' operations. ICANN's role vis-a-vis these new top level domains is historically without precedent. It is dramatically different from the role played by Jon Postel, who was largely responsible for the governance of the domain name system until his death in 1998. Yet while ICANN's activities are unlike Postel's, they are unexpectedly familiar to the United States communications lawyer: ICANN's actions strikingly parallel the Federal Communications Commission's historic course in licensing broadcasters.[1]

ICANN has selected top level domain registries through processes that, if they were vastly improved, would look like the FCC's historic "public interest"-based comparative licensing. Like the FCC, ICANN has used this licensing process to regulate registry conduct, although ICANN's regulation goes far beyond anything the FCC ever attempted. And as with the FCC, ICANN's regulatory imperative has flowed largely from scarcity – in this case, the scarcity of generic top level domains in the ICANN root. The scarcity of top level domains is not a technological artifact, though, as with broadcast licensing; rather, ICANN is maintaining it as a policy matter.

This paper provides a history: It tells the story leading to ICANN's selection of seven new top-level domains a year-and-a-half ago. In telling that story, and selecting from the universe of facts to include, I will focus on facts illuminating two basic themes. The first of those themes relates to the method that ICANN chose to select the new TLD registries. ICANN's selection process was badly dysfunctional; it was described by one media observer as "torturous," "channeling the spirit of [Walt] Disney," "a bad parody of Florida's election process," and "bizarre."[2] ICANN's incoming chairman compared the selection process to that of a venture capital firm, and urged that ICANN find a way to "extract" itself.[3] How did ICANN reach that point, and what alternatives did it have? What alternatives, indeed, does it have now?

---

[*] Professor of Law, Wayne State University. I owe thanks, as always, to Jessica Litman; without her, this narrative would be even more turgid, and would have rather less to say. I was the co-chair of a working group established as part of the ICANN process to formulate recommendations regarding the deployment of new generic top-level domains. The view expressed in this article, however, are solely my own.

[1] In the words of Harold Feld: "ICANN recapitulates the FCC, and does it badly." Email message from Harold Feld, Associate Director, Media Access Project, to the author (Feb. 21, 2001).

The second theme relates to the degree of ICANN's control over the day-to-day operations of the new registries. After ICANN's selection of the seven new registries, the registries and ICANN staff sat down to negotiate contracts. ICANN staff had originally contemplated that negotiating all the contracts would take no more than six weeks.[4] Instead, as of this writing (ten months after the registries were selected), ICANN has completed agreements with only three, and negotiations with the other four are still ongoing. The most important reason for this delay is the extraordinarily detailed and comprehensive nature of the new contracts; a single one is about two inches thick in hard copy. The contracts incorporate an extensive set of commitments by the registries to ICANN, with ICANN specifying many aspects of their operations; their negotiation amounts to extensive regulation on ICANN's part of registry activities. What led ICANN to seek to impose that regulation, and is it necessary?

I will not resolve these questions in this paper; I will address the first to a limited extent, and the second not at all. I will leave the answers to a later, longer, article. What I am presenting in this paper, rather, is simply a history. But it may be useful to the reader, in reading that history, to keep these questions in mind.

## I. BACKGROUND

---

[2] Brock Meeks, ICANN and the seven dwarves (Nov. 22, 2000), <http://www.msnbc.com/news/493721.asp>.

[3] Vinton Cerf, Chairman, ICANN, quoted in id.

[4] See, e.g., New TLD Application Process Overview (Aug. 3, 2000), <http://www.icann.org/tlds/application-process-03aug00.htm> (explaining that the Board would select the new registries in mid-November, and setting December 31 as the target date for completing negotiations).

A. *Technical basis of the DNS*

The domain name system matches *Internet protocol (IP) addresses*, which identify individual host computers on the Internet, with *domain names*. An IP address is a unique 32-bit number, usually printed in dotted decimal form, such as 128.127.50.224;[5] a domain name is a set of text labels set off by dots, such as threecats.net or law.wayne.edu.[6] A system matching names to numbers, so that a user can locate an Internet resource knowing only its domain name, has two advantages. First, domain names are relatively easy to remember and to type. IP addresses, by contrast, are opaque and harder to remember. Second, the use of domain names provides a "level of indirection" making it possible for network operators to change the IP addresses associated with various machines while leaving their names – which outsiders use to find them – untouched.[7]

The current domain name system, developed by Postel and Paul Mockapetris (both of the University of Southern California's Information Sciences Institute), is hierarchical.[8] The overall name space is divided into top level domains, or TLDs; each top-level domain is divided into second level domains. At each level, the pyramidal structure of the name space replicates itself. The owner of each second level domain is at the apex of a pyramid consisting of the third level domains (if any) within that second-level domain, and so on.[9]  Thus, the .edu TLD is divided into about 4000 second level domains such as wayne.edu; wayne.edu is divided into third level domains including law.wayne.edu, gradschool.wayne.edu, and socialwork.wayne.edu.

---

[5]     My reference to an IP address as "unique" is oversimplified. Techniques such as network address translation can allow a computer to function using an IP address that is unique only within that computer's local network. See Jonathan Weinberg, Hardware-Based ID, Rights Management, and Trusted Systems, 52 Stan L. Rev. 1251, 1260 n. 22 (2000). Most residential Internet users get their IP addresses pursuant to a dynamic allocation system under which the user may get a different address each time she logs on to her Internet service provider. See id. at 1260.

[6]     "A name indicates what we seek. An address indicates where it is." Jon Postel, RFC 791, DARPA Internet Program Protocol Specification, <http://www.rfc-editor.org/rfc/rfc791/txt>(1981).

[7]     See Ted Byfield, DNS: A Short History and a Short Future, 4 First Monday 3, para. 34 (Mar. 1, 1999), <http://www.firstmonday.dk/issues/issue4_3/byfield/index.html>.

[8]     See Paul Mockapetris, RFC 882, Domain Names – Concepts and Facilities, <http://www.rfc-editor.org/rfc/rfc882/txt>(1983).

This hierarchy makes it easy for the job of name-to-number translation to be shared by a large number of servers.  At the apex of the DNS pyramid is a set of thirteen root servers, each of which lists the IP addresses of the computers containing the zone files for each of the top-level domains.  At the next level are the computers holding those top-level domain zone files, each of which lists the IP addresses of the name servers for each second-level domain it controls, and so on.  When a user looking for a particular Internet resource types in a domain name, her computer begins at the bottom of the pyramid: it queries a set of local DNS servers, specified in its software, to find the IP address corresponding to that domain name.  If those local servers do not know the answer, they move the request up the line.

This structure has far-reaching implications.  On the one hand, it lends itself to decentralization, since the person controlling any given host can adopt policies governing registration below it (but not elsewhere) in the pyramid.  The owners of wayne.edu, for example, have complete control over who they will allow to register third-level domains such as law.wayne.edu; there is no snorlax.wayne.edu, because that label does not fit within the naming scheme that the proprietors of wayne.edu established.

On the other hand, control over the root zone -- at the very top of the pyramid -- carries with it considerable power.  If a user types in a domain name incorporating a top-level domain that is unknown to the root servers, then the DNS will be unable to find the corresponding computer.  The power to control the root servers, thus, is the power to decide (1) which top-level domains are visible in the name space; and (2) which name servers are authoritative for those top-level domains – that is, which registries get to allocate names within each of those top-level domains.  Historically, the Internet root zone was overseen by Postel and others at USC's Information Sciences Institute; beginning in the late 1980s, their activities coordinating the root

---

zone and IP address allocation came to be referred to as the Internet Assigned Numbers Authority (IANA).[10]

There is no technical or legal requirement that a person use the root servers established by IANA to resolve DNS queries. Users can point their computers at entirely different DNS servers that in turn point to different root servers, referencing a different set of top-level domains.[11] Such alternative root servers do exist, so that if one points one's computer at the right DNS server, one can send email to addresses that the rest of the Internet does not recognize, such as <richard@vrx.zoo>.[12] Very few Internet users, though, look to alternative root servers. The vast majority rely on the single set of authoritative root servers, historically supervised by Postel and IANA, that have achieved canonical status.[13]

B. *Building the domain name space*

The first top level domains set up in the current domain name system, beginning in January 1985, were .arpa (which during an initial transitional period contained all then-existing Internet hosts,[14] and now is limited to certain infrastructural functions); .com (initially intended for businesses);[15] .edu (for universities);[16] .gov (for U.S. government agencies), .mil (for the U.S. military); .net (for Internet "network-type organizations," such as network service centers and consortia or network information and operations centers);[17] and .org (for entities "that do not clearly fall within the other top-level domains").[18] Only one other of these so-called generic domains was created during Postel's lifetime – the .int domain, for international treaty organizations, in 1988.[19]

Beginning in February 1985, though, Internet engineers began adding "country-code" top level domains (ccTLDs) to the root zone.[20]  The first ones added were .us, for the United States; .gb and .uk, for Great Britain; .il, for Israel, and .au, for Australia.[21]  Early in 1994, Postel memorialized the criteria for adding new top-level domains in a document known as *RFC 1591 (Domain Name System Structure and Delegation).*[22]  At the time, he was adding new country-code TLDs at a rate of about one every sixteen days;[23] he had created more than one hundred since 1985.

---

[15]     See Mary Stahl, RFC 1032, Domain Administrators Guide, <ftp://ftp.isi.edu/in-notes/rfc1032.txt> (1987).

[16]     Id.

[17]     Id.

[18]     Id.

[19]     See email from Anthony M. Rutkowski, Vice-President, Internet Strategies, Verisign-NSI, to the author (July 24, 2000) [hereafter, Rutkowski email].

[20]     The criteria for adding new ccTLDs, in those early days, were arrived at by discussion between Postel and others at the Information Sciences Institute, and Mary Stahl and others at the Network Information Center then operated by SRI International.  See email from Mary Stahl, formerly hostmaster, SRI-NIC, to the author (Aug. 14, 2001); email from Anthony M. Rutkowski, Vice-President, Internet Strategies, Verisign-NSI, to the author (July 26, 2001).  As a general matter, those ccTLDs utilized the services of researchers with ties to the U.S. Defense Department.  Email from Mary Stahl, supra.

[21]     See Rutkowski email, supra n. 19.

[22]     Postel, supra n. 9.  This document explains the criteria for IANA's "coordination and management of the Domain Name System (DNS), and especially the delegation of portions of the name space called top-level domains."  Id.

[23]     See Rutkowski email, supra n. 19 (23 new TLDs were added in 1993, and 22 in 1994).

Before Postel would add a new country-code top level domain, the following requirements had to be met.  First, "significantly interested parties" within the country in question had to agree on a manager to supervise the domain.  Postel emphasized that the burden was on contending parties within a country to reach agreement among themselves; he would not change a delegation once made, absent substantial misbehavior by the manager,  unless all of the contending parties agreed on the change.[24]  Second, the proposed ccTLD manager had to understand its responsibilities.  A ccTLD manager, RFC 1591 emphasized, is a trustee for the people of the nation in question, with a duty to serve the local community, and a trustee for the global Internet community as well.  It must operate the domain in a technically competent manner, maintaining  adequate Internet connectivity.  It must treat all users equally, processing requests in a non-discriminatory fashion, and treating academic and commercial users on an equal basis. [25]

Apart from these general considerations, though, RFC 1591 conspicuously avoided any instructions about how a new country-code domain should be run.  RFC 1591 said nothing further about a registry's business model.  It did not speak to whether a registry should charge for domain name services, or whether it should limit registration to residents of the country in question.  It said nothing about how the registry should structure the name space within the ccTLD.  Indeed, it said very little about the registry's technical operations.  These decisions were up to the manager of the domain; they were no business of IANA's.[26]

---

[24]　　ICANN has since departed from this rule.  In late August 2001, ICANN (wearing its IANA hat) redelegated the .AU ccTLD (for Australia) from Robert Elz to the newly formed Australian Domain Name Administration, over Elz's objection.  See IANA Report on Request for Redelegation of the .au Top-Level Domain (Aug. 31, 2001), <http://www.iana.org/reports/au-report-31aug01.htm>; A. Michael Froomkin, How ICANN Policy Is Made (II) (Sept. 5, 2001),<http://www.icannwatch.org/essays/dotau.htm>.

[25]　　Postel, supra n. 9, sec. 3.

[26]　　RFC 1591 did say one further thing relating to how a ccTLD registry should be run: It noted that the registry had "no role or responsibility," other than providing appropriate contact information to the parties, when a trademark owner challenged a domain name holder's right to the name.  That is, the registry was not to provide an extrajudicial remedy to a trademark owner unhappy with the registration of a particular name.  If a domain name registrant was violating trademark rights, that was the registrant's own legal concern, but it was no concern of the operator of the domain.  See id. at sec. 4.

In RFC 1591, Postel stated that it was "extremely unlikely" that any new generic TLDs would be created.[27] In the mid-1990s, though, dissatisfaction with the domain name system began to mount. Registration services in .com, .net, .org and .edu were then performed by a company known as Network Solutions, Inc (NSI), pursuant to a cooperative agreement with the U.S. National Science Foundation (NSF). Initially, NSF paid for all registrations, which were free to users;[28] as the number of registrations began to rise, though, NSF and NSI agreed to take the U.S. government out of the funding loop. Rather, NSI would charge a $50 annual fee to each domain name registrant.[29]

The NSI fee crystallized growing unhappiness with the structure of the domain name system. Registrants wondered why, in seeking to register names in the generic top level domains, they were stuck with the service provided, and the fees charged, by the NSI monopoly.[30] NSI also generated animosity with its domain name dispute policies, under which it would suspend a domain name upon receiving a complaint from a trademark owner, without regard to whether the trademark owner had a superior legal claim to the name.[31] Many saw the dominance of the .com domain in the name space as unhealthy.[32] Finally, there was growing consensus in the technical community that the architecture would support many more top-level

---

[27]     Id. sec. 2.

[28]     See NSF Cooperative Agreement No. NCR-9218742, <http://www.networksolutions.com/en_US/legal/internic/cooperative-agreement/agreement.html> (Jan. 1, 1993).

[29]     See NSF Cooperative Agreement No. NCR-9218742, Amendment 4, <http://www.networksolutions.com/en_US/legal/internic/cooperative-agreement/amendment4.html> (Sept. 13, 1995).

[30]     See email from Jon Postel to Rick Adams, Chairman and Chief Technical Officer, UUNET (Sept. 15, 1995) ("I think this introduction of charging by [NSI] for domain registrations is sufficient cause to take steps to set up a small number of alternate top level domains managed by other registration centers. I'd like to see some competition between registration services to encourage good service at low prices.").

[31]     See Ellen Rony & Peter Rony, The Domain Name Handbook 147 (1998); Carl Oppedahl, Avoiding the Traps in the New Rules for Registering a Domain Name, N.Y.L.J., Aug. 8, 1995, at 5.

[32]     See Jon Postel, New Registries and the Delegation of International Top-level Domains, draft-postel-iana-itld-admin-02.txt (Aug. 1996), <http://sunsite.org.uk/rfc/draft-postel-iana-itld-admin-02.txt>, at sec. 1.5.2 ("the inherent perceived value of being registered under a single top level domain (.COM) is undesirable and should be changed").

domains than had been authorized so far.[33]

Accordingly, in 1996, Postel suggested that IANA authorize up to 150 new generic top-level domains to be operated by new registries.[34]  The qualifications he deemed necessary for a person or organization seeking to operate one of the new domains were lightweight.  First, the applicant would have to show that it could provide a minimum set of registration services: maintenance of up-to-date registration data in escrowable form, capability to search the second level domain database via the whois protocol, live customer support during business hours, etc.[35]  Second, it would need adequate Internet connectivity, and at least two nameservers in geographically diverse locations running an up-to-date version of the BIND software.[36]  Finally, it would need to present some documentation lending credibility to the conclusion that it was proposing a viable business, "likely to operate successfully for at least five years."[37]

Postel was emphatic, though, that a person applying to operate a new gTLD would not have to submit a business plan, and that "[i]nternal database and operational issues . . . including pricing to customers of the registry" were no business of IANA's.  These were "free-market issues," to be decided by each registry for itself.[38]

---

[33]　See id. at secs. 4 ("further growth within the iTLDs can be accommodated technically"), 1.5.2 (while it would be administratively burdensome to have "enormous numbers (100,000+)" of top level domains, diversity in the top-level domain space is the best way to ensure quality service).

[34]　Id. at secs. 5.6, 6.1.

[35]　Id. sec. 6.4.1.

[36]　Id. sec. 6.4.2.

[37]　Id. sec. 6.4.3.  Each applicant would submit a statement of its proposed charter, policies, and procedures; a statement of how one would qualify to register in the proposed TLD; a commitment to treat applicants in a non-discriminatory manner; and a statement demonstrating its organizational and technical competence. Id. sec. 6.5.3.  An ad hoc committee established by IANA would then review the application according to the criteria set out above.  See id. sec. A.3.  The committee would also have the job of devising a selection method where multiple applicants sought to run the same TLD.  See id. sec. 6.7.

[38]　Id. secs. 6.4.1, 6.4.3.  The document at a later point, though, does caution that a registry's pricing should be "within the bounds of the policy [it] published when it was chartered."  Id. sec. 8.1.

Postel's proposal met with a guardedly favorable reaction from the Internet Society[39] (a nonprofit membership organization that is home to key Internet technical bodies).[40] Other groups, however, soon came forward to object. Postel's plan only began a long and contentious process in which participants debated the nature of new TLDs and the future of Internet governance. That story has been told elsewhere;[41] let it suffice that two years later the U.S. government determined that "the challenge of deciding policy for new domains" should be put in the hands of a new nonprofit corporation that would step into IANA's shoes.[42]

Historically, all of the major actors involved with the name space had fulfilled their responsibilities pursuant to agreements with the U.S. government. USC's Information Sciences Institute, which housed Postel, had long had contracts with the U.S. Defense Department covering the IANA work; NSI, which operated the registry for the .com, .net., .org and .edu domains, did so pursuant to a cooperative agreement with the National Science Foundation. As part of its solution to the controversies raging over the domain name space, the U.S. government determined that it should "withdraw from its existing management role" in favor of a new, not-for-profit corporation formed and run by "private sector Internet stakeholders."[43] The new corporation, which would manage domain names, the IP address allocation system, and the root server network, would be run by a board of directors broadly reflecting the Internet private sector. The U.S. government would recognize it by entering into agreements with it that would give it effective policy authority over the root zone.

---

[39]     See Internet Society, Minutes of the Annual General Meeting of the Board of Trustees (June 24-25, 1996), <http://www.isoc.org/isoc/general/trustees/mtg09.shtml>, item 8.3.

[40]     See All about ISOC, <http://isoc.org/isoc>.

[41]     See, e.g., A. Michael Froomkin, Wrong Turn in Cyberspace: Using ICANN to Route Around the APA and the Constitution, 50 Duke L.J. 17 (2000); Milton Mueller, ICANN and Internet Governance: Sorting Through the Debris of 'Self-Regulation", 1 Info 497 (1999); Simon, supra n. 13; Jonathan Weinberg, ICANN and the Problem of Legitimacy, 50 Duke L.J. 187 (2000).

[42]     Management of Internet Names and Addresses, 63 Fed. Reg. 31,741, 31,746 (1988) [hereafter, White Paper].

[43]     Id. at 31,749. This aspect of the initiative followed the Clinton Administration's motto that in matters affecting electronic commerce, "[t]he private sector should lead." The White House, A Framework for Global Electronic Commerce (July 1, 1997), <http://www.ecommerce.gov/framewrk.htm>.

In late 1998, after an extended series of negotiations between IANA and NSI – and consultations with the U.S. government, a variety of foreign governments, large corporations, and others – Postel took a crucial step to implement the White Paper's direction by transmitting to the U.S. Department of Commerce documents creating the new corporation.[44] These documents included the articles of incorporation of the new Internet Corporation for Assigned Names and Numbers; biographies of a proposed initial board of directors; and a set of proposed bylaws. The new directors were drawn, for the most part, from the worlds of telecommunications and information technology; few of them had specialized knowledge of the Internet or of domain name issues.[45] The plan was that the board members would be guided by the wisdom of Postel as the new corporation's chief technical officer and could lend their influence and neutrality to bolster his decisions.[46]

Two weeks later, Jon Postel died of complications following open heart surgery. This was a tremendous blow to the new organization; on what basis, now, were industry members and the public to have faith in ICANN's decision-making? The U.S. government, though, had issued its policy statement and committed itself to the new organization. It pushed forward. It solicited public comment on ICANN's proposal, and began negotiating with ICANN's lawyer (Joe Sims of the Jones, Day law firm) over failings in the proposed bylaws. Ultimately, the government entered into a memorandum of understanding with ICANN, recognizing it and authorizing it to exercise DNS management functions subject to the government's continuing oversight.

ICANN came into existence under a cloud. Its board members, who had been chosen in a closed process, were many of them unknown to the Internet community. While ICANN had the U.S. government's seal of approval, the government's own authority over the DNS was murky and contested.[47]  There were some who contended that ICANN was simply illegitimate.[48]  On the other hand, ICANN had control of several of the levers of power.  Most importantly, with the U.S. government's support, it had policy control of the root zone, because NSI operated the primary root server subject to U.S. government instructions.[49]  The U.S. government, moreover, was able to use its negotiating leverage to cause NSI to recognize ICANN's policy authority (while NSI simultaneously secured favorable terms for itself relating to its ability to exploit the lucrative .com, .net and .org top level domains).  Finally, ICANN was tasked by the Department of Commerce with supervising a process under which multiple new competitive "registrars" would sell domain names in the NSI-operated TLDs.  Any company wishing accreditation as a registrar, therefore, had to recognize ICANN's authority and agree to its terms.

---

[44] See letter from Jon Postel, Director, IANA, to William Daley, Secretary of Commerce (Oct. 2, 1998), <www.ntia.doc.gov/ntiahome/domainname/proposals/icann/letter.html>.

[45] See Froomkin, supra n. 41, at 73; Weinberg, supra n. 41, at 209-10. Rather, the organizers hoped for a board of directors that was not tainted by previous involvement in the DNS wars. As Postel put it, the goal in selecting a board was "to see that policies and procedures are developed in a fair and open manner, not to rehash all the arguments and positions that go into developing those policies and procedures." Email message from Jon Postel to the IETF mailing list (Sept. 27, 1998) (on file with author).

[46] See Froomkin, supra n. 41, at 72.

The new organization's internal structure was complex.  In theory, the job of developing policy was lodged in three "Supporting Organizations" – one to address policy relating to domain names, one for policy relating to IP address allocation, and one for policy relating to "the assignment of parameters for Internet protocols."[50]  The organization charged with developing policy relating to domain name issues was the Domain Name Supporting Organization (DNSO); within that body, policy authority was exercised by a Names Council, whose membership was selected by seven industry groupings (known in ICANN lingo as "constituencies").[51]

According to ICANN's bylaws, the Names Council has "primary responsibility for developing" domain name policy within the ICANN structure.[52] It is supposed to do this by managing a "consensus building process" within the DNSO; it has the power to designate committees and working groups to carry out its substantive work.[53] If the Names Council determines that the DNSO has produced a "community consensus" on some matter of domain name policy, it is to forward that consensus to the Board.[54] The bylaws state that as a general matter, ICANN may not enact domain-name policy without the approval of a Names Council majority.[55]

---

[47]      See Weinberg, supra note 41, at 204-05.

[48]      See id. at 213.

[49]      ICANN also needed the cooperation of the operators of the other root servers.  Those players, however, were loathe to introduce domain name resolution conflicts by failing to faithfully mirror the root zone as displayed in NSI's primary root server, and – correctly – saw ICANN as having more legitimacy than any competitor.

[50]      In fact, ICANN has no meaningful policymaking responsibilities relating to protocol parameter assignment.  The "protocol" supporting organization was created, rather, to assure a voice for the Internet engineering community in ICANN decision-making.  See Weinberg, supra note 41, at 236-337.

[51]      See id. at 238-42; Jonathan Weinberg, Geeks and Greeks, 3 Info 313, 327-28 (2001).

These formal rules, though, grossly misdescribe the actual ICANN process. The Names Council has turned out to be incapable of generating detailed policy recommendations, and the DNSO has not proved to be an important locus for policy development.[56] Rather, that role has been taken over by ICANN staff.[57]

## II. ADDING NEW TOP LEVEL DOMAINS

When ICANN was formed, the most important substantive policy question facing the new organization was whether, and under what circumstances, it would add new generic top level domains to the name space. On May 27, 1999, ICANN's board of directors instructed the DNSO to formulate recommendations on the question of adding new generic top level domains.[58] The DNSO in turn passed the matter to a working group.

---

[52]     ICANN Bylaws, art. VI(2)(b), <http://www.icann.org/general/bylaws.htm#VI>

[53]     ICANN Bylaws, art. VI-B(2)(b), <http://www.icann.org/general/bylaws.htm#VI-B>.

[54]     Id. art. VI-B(2)(d).

[55]     See id.; id. art. VI(2), <http://www.icann.org/general/bylaws.htm#VI>.

[56]     See Jonathan Weinberg, Review of the Domain Name Supporting Organization (Nov. 11, 2000), <http://www.law.wayne.edu/weinberg/dnso_review.htm>.

[57]      During the time period covered by this paper, key ICANN staff personnel were CEO Mike Roberts, Chief Policy Officer Andrew McLaughlin, and Vice President & General Counsel Louis Touton. Because the documents they issued on behalf of the corporation were usually unsigned, so that their individual roles could not be discerned, I will refer to them collectively here as "staff."

By now, it had become clear that Postel's proposal to add hundreds of new top level domains, although technically straightforward, was politically infeasible. Trademark lawyers had organized early to oppose any expansion of the name space. They feared that increasing the number of TLDs would force trademark owners, seeking to prevent the registration of domain names similar or identical to their trademarks, to incur higher policing costs.[59] At the very least, the trademark bar argued, before there could be any expansion of the name space there had to be a well-established, thoroughly tested mechanism built into the DNS architecture that would allow trademark owners to gain control of offending domains without going to court.[60] The trademark lawyers convinced leaders of the technical community that they had the political clout to stop any expansion of the name space to which they had not agreed.

In the DNSO's working group, the battles raged anew. Some participants repeated that ICANN should immediately add hundreds of new top-level domains.[61] Such a step would maximize consumer choice, making many new appealing names available. It would ensure meaningful competition among top level domain registries, eliminating market-power problems that were unavoidable with a smaller number. It would minimize trademark problems, because consumers, understanding that a given SLD string could belong to different registrants in different TLDs, would not be confused into thinking that any given domain name was associated with a given company.

---

[58] See Minutes: Meeting of the Initial Board (May 27, 1999), <http://www.icann.org/minutes/minutes-27may99.htm> (Resolution 99.48).

[59] See GTLD-MoU Frequently Asked Questions – Why did you choose these names and why only seven?, <http://www.gtld-mou.org/docs/faq.html#2.2> (noting the "adamant" position of the trademark bar in 1996-97 against name space expansion). In an elaborate Internet Ad Hoc Committee (IAHC) that had been formed by IANA and the Internet Society prior to ICANN's creation, the representative of the International Trademark Association had urged that there should no expansion of the top level domain space at all. Email from Dave Crocker to the domain-policy mailing list (Aug. 7, 2000) (on file with author). Largely because of trademark opposition, the IAHC ended up recommending the addition of just seven new top level domains. For background on the IAHC, see Simon, supra n. 13; Final Report of the International Ad Hoc Committee: Recommendations for Administration and Management of gTLDs (1997), <http://www.gtld-mou.org/draft-iahc-recommend-00.html>.

[60] See GTLD-MoU Frequently Asked Questions – Why did you choose these names and why only seven?, <http://www.gtld-mou.org/docs/faq.html#2.2>.

[61] See Interim Report of Working Group C of the Domain Name Supporting Organization, Internet Corporation for Assigned Names and Numbers, Position Paper B (Oct. 23, 1999), <http://www.dnso.org/dnso/notes/19991023.NCwgc-report.html#Position Paper B>.

Trademark lawyers, by contrast, urged that no new TLDs should be added until a set of new trademark protections had been built into the system and it was "clear that the proposed safeguards are working"; only then, the opponents indicated, would they entertain the possibility of introducing one or more new gTLDs "on an as needed basis."[62]  Nor were trademark lawyers the only group expressing skepticism about expanding the name space.  Business players that had prospered under the existing system worried about disruptive change.[63]  Internet service providers worried that name space expansion would encourage their users to acquire their own domain names, weakening the link between user and ISP and increasing the ISP's costs. Existing commercial domain name registries (NSI and a few of the ccTLDs) saw new top level domain registries as competition.

After extensive debate, the working group reached what it termed "rough consensus" (defined as a two-thirds vote of its members) in support of a compromise position, put forward by the group's co-chair, under which ICANN would begin by adding six to ten new gTLDs, followed by an evaluation period.[64]  It agreed as well that the initial rollout should include a wide range of top level domains, including both "open" TLDs, in which anyone could register, and restricted TLDs for the benefit of particular groups.[65]

---

[62]    See Interim Report of Working Group C of the Domain Name Supporting Organization, Internet Corporation for Assigned Names and Numbers, Position Paper C (Oct. 23, 1999), <http://www.dnso.org/dnso/notes/19991023.NCwgc-report.html#Position Paper C>.

[63]    See, e.g., John C. Lewis, Business & Commercial Constituency of the DNSO Submission on the Creation of New gTLDs , email message from John C. Lewis, Manager - International Organizations Europe, British Telecom, to the author (Jan. 10, 2000).

[64]    Report (Part One) of Working Group C (New gTLDs) (Mar. 21, 2000), <http://www.icann.org/dnso/wgc-report-21mar00.htm>.  For the most part, proponents of a much faster rollout nonetheless agreed to support this position as a compromise; proponents of much more limited expansion voted against it.  Full disclosure: I was the working group's co-chair, and the author of the compromise proposal.

[65]    See Supplemental Report to Names Council Concerning Working Group C (Apr. 17, 2000), <http://www.icann.org/dnso/wgc-supp-report-17apr00.htm>.

But the working group failed to reach consensus on other issues. Some within the working group had urged that ICANN should require all registries in the initial rollout to be operated on a not-for-profit, cost-recovery basis;[66] others argued, just as strongly, for a mix of for-profit and not-for-profit registries. The working group was able to come to no resolution on this point.[67] More importantly, the working group failed to resolve *how* ICANN should select the new top-level domains.[68]

The Names Council, upon receiving the working group report, declined to fill in the gaps. It agreed on a general statement supporting the introduction of new gTLDs but recommending that their introduction be "measured and responsible," giving due regard to the goals of generating an "orderly" process for initial registration in the new domains; protecting intellectual property rights; and safeguarding user confidence in the technical operation of the domain name space.[69] The Names Council statement said little about the number of new gTLDs, the nature of the new registries, or how they should be selected.[70]

This left ICANN staff, tasked by the Board with bringing new gTLDs online, with freedom of action. After the Names Council pronouncement, ICANN staff released a report, styled a "discussion document," stating that the addition of new top level domains should be well-controlled and small-scale, with the goal of establishing a "'proof of concept'" for possible future introductions – that is, that the point of the initial rollout would simply be to establish (or

---

[66]     The proponents of the nonprofit-registries-only approach had earlier been supporters of the IAHC plan, under which all new gTLDs would be operated by a single not-for-profit registry, organized as a cooperative venture of participating registrars. See supra n. 59.

[67]     See email message from Jonathan Weinberg, co-chair, working group C to <wg-c@dnso.org> (Sept. 3, 1999).

[68]     See email message from Jonathan Weinberg, co-chair, working group C to <wg-c@dnso.org> (Mar. 16, 2000).

[69]     DNSO Names Council Statement on new gTLDs (April 19, 2000), <http://www.dnso.org/dnso/notes/20000419.NCgtlds-statement.html>.

[70]     The Names Council did suggest that "[t]o assist the Board in the task of introducing new gTLDs," ICANN staff should "invite expressions of interest from parties seeking to operate any new gTLD registry, with an indication as to how they propose to ensure to promote these values." Id.

disprove) the proposition that new top-level domains *could* be added to the name space successfully.[71]

The report requested public comment on seventy-four policy and technical questions.[72] There were a variety of questions, though, that the document did *not* ask. The document elaborately justified, and treated as settled, its conclusion that any introduction of gTLDs should be small-scale, intended only to serve as "proof of concept." The "proof of concept" notion was not intuitively obvious, since it was not entirely clear what concept was to be proved: It was already abundantly clear that adding new gTLDs was *technically* feasible, and would not threaten successful name resolution. After all, IANA had added ccTLDs to the root zone quite frequently over the years, and adding a gTLD was no different from adding a ccTLD from the standpoint of whether domain name servers would return accurate responses to DNS queries.[73] But staff was on relatively firm ground in calling for a small-scale rollout: The Names Council had requested "measured and responsible" introduction, and had noted its concern that the introduction of a large new gTLD would be marred by lack of "orderly" process and developments unfavorable to trademark owners. The DNSO's working group, along similar lines, had suggested that the initial rollout of six to ten be followed by "evaluation" before ICANN proceeded further.

---

[71] ICANN Yokohama Meeting Topic: Introduction of New Top-Level Domains (June 13, 2000), <http://www.icann.org/yokohama/new-tld-topic.htm>.

[72] They covered such matters as how to address "stability concerns" in an initial rollout; what lessons ICANN should seek to learn in that introduction; whether the rollout should be directed to securing competition among registries hosting open top level domains, and if so how; whether it should be directed toward increasing the utility of the domain name space as a resource-locating tool, through the introduction of limited-purpose top level domains, and if so how; who should formulate policy for limited-purpose TLDs; what additional privileges ICANN should give trademark owners in connection with, or as a prerequisite for, the addition of new gTLDs; on what schedule ICANN should seek to select the new TLD registries; and what information (such as proposed TLD string, nature of the proposed TLD, justification for the proposed TLD, financial data, proposed business model, technical capabilities, mechanisms proposed to benefit trademark owners, etc.) each applicant should have to provide. Id.

[73] The concept being proved, thus, was more amorphous: Vint Cerf, ICANN's current chair, has characterized it as whether "it is possible to introduce new top-level domains in the DNS at a time when the economic importance/value of domains is very different from the time when the first gTLDs were created." We did not know, he suggests, "what it means to operate TLDs in the rapidly evolving commercial context of today's (and tomorrow's) Internet." Vint Cerf Replies to (most of) Your Questions (Apr. 19, 2000), <http://www.icannwatch.org/article.php?sid=114>.

Also implicit in the staff document was its rejection of any suggestion that new top level domain registries had to be not-for-profit. The discussion document assumed that the new TLDs would be run by multiple new entities that had applied to ICANN for the right to do so, and it explicitly contemplated that at least some of those registries would be profit-oriented firms.[74] The report contained no discussion recognizing that these were, in fact, decisions.

Most important were choices about how the new registries would be chosen. When ICANN inserts a new TLD into the root, the new zone file entry reflects a series of choices. The zone file must identify the string of letters that will sit at the right of all domain names in the new TLD, such as ".edu" or ".info". It must also identify the particular organization that will administer the master registry database for that TLD, and enter the IP addresses of name servers controlled by that organization into the root zone. In considering how ICANN should go about selecting new TLDs, the DNSO's working group had confronted a range of options. Should ICANN first identify the TLD strings that would be desirable additions to the name space, identify how it wanted those TLDs to be run, and only then solicit applications for registries to operate the TLDs according to its specifications? If so, should it establish a master plan (such as a Yellow Pages-style taxonomy), or should it identify the desirable new TLD strings on an ad hoc basis? Or should ICANN take an alternative approach, picking a set of registries according to objective criteria, and afterwards allowing the selected registries to choose their own strings? Or should each would-be registry apply to ICANN, explaining which string or strings it wished to run a registry for, so that ICANN could select registry and string together?[75]

The staff report answered all of these questions: It charted a path in which ICANN would request an application from each organization seeking to operate a new gTLD. Each of these organizations would set out its business, financial and technical qualifications, together with the

---

mechanisms it proposed for the benefit of trademark owners, its proposed TLD string, and the characteristics of the proposed top level domain. It would address such issues as the market targeted by the proposed TLD, and the TLD's criteria for registration. The staff document thus eliminated at the outset such possibilities as first identifying the TLD strings that would be desirable additions to the name space and only then soliciting applications for registries to operate the TLDs in question, or picking a set of registries according to hard-edged, objective criteria, without regard to the nature of the TLDs they wished to run. Rather, the document -- essentially without discussion of alternatives[76] -- assumed a process in which ICANN, picking a small number of TLDs to allow into the initial rollout, would look at all relevant aspects of every proposal and decide which ones presented the best overall combination of TLD string, TLD charter, business plan, robust capitalization, and other (incommensurable) factors.

When staff made this choice, some aspects of the resulting process were predictable. Anyone familiar with the Federal Communications Commission (FCC) comparative hearing process for broadcast licenses can attest that this sort of ad hoc comparison is necessarily subjective.[77] Before the fact, it is difficult to predict what results such a process will generate; afterwards, it is hard to justify why one proposal was chosen and not another. Because decisions are unconstrained by clear-cut rules, the process lends itself to arbitrariness and biased application. Yet the process had advantages that appealed to ICANN decision-makers. The Board members, in comparing the applications, would be free to take their best shots, in a situationally sensitive manner, at advancing the policies they thought important. They would not have to worry about being bound by hard-and-fast rules yielding unfortunate results in the particular case.[78]  More importantly, given business and trademark lobbyists' fear of new gTLDs and their potential for disruptive change, this approach allowed ICANN to maintain the greatest degree of control over the selection process.[79]  It gave assurance that any new gTLDs emerging from the process would be not only few, but also safe.

---

[76] The document did at one point request comment on this issue – "[s]hould ICANN select the TLD labels, should they be proposed by the applicants for new TLD registries, or should they be chosen by a consultative process between the applicants and ICANN?" ICANN Yokohama Meeting Topic: Introduction of New Top-Level Domains (June 13, 2000), <http://www.icann.org/yokohama/new-tld-topic.htm>, at sec. IV.1 (Q54). This question, though, was isolated; the discussion reflecting the document's answer was pervasive

[77] See Jonathan Weinberg, Broadcasting and Speech, 81 Calif. L. Rev. 1101, 1168-69 (1993). In terminology common in legal philosophy, the process relies on "standards" rather than "rules." See id. at 1167-69; Duncan Kennedy, Form and Substance in Private Law Adjudication, 89 Harv. L. Rev. 1685, 1685, 1687-98 (1976); Kathleen M. Sullivan, The Supreme Court, 1991 Term–Forward: The Justices of Rules and Standards, 106 Harv. L. Rev. 22, 58-59 (1992).

ICANN's board of directors formally authorized submission of applications to operate the new TLDs,[80] and staff published a remarkably detailed application form.[81]  ICANN instructed prospective registry operators that they had to complete and return the forms in six weeks.[82] Each application was to be accompanied by a non-refundable $50,000 fee,[83] to cover the costs of what staff described as a "very intensive review and analysis of applications on many levels

---

[78]     See Weinberg, supra note 77, at 1168-69.

[79]     I must confess that back in October 1999, I suggested adoption of an ad hoc process on precisely this ground – that it would "would likely make ICANN itself most comfortable," and "as a matter of supervising the initial rollout, it would be responsive to oft-expressed concerns about Internet stability and reliability."  I cautioned, though, that in the long term "such an approach would not be desirable; it presents the risk of subjective and unaccountable decision-making."  Interim Report of Working Group C of the DNSO, Position Paper A, <http://www.dnso.org/dnso/notes/19991023.NCwgc-report.html#Position Paper A>.

[80]     Resolutions of the ICANN Board on New TLDs (July 6, 2000), <http://www.icann.org/tlds/new-tld-resolutions-16jul00.htm>.

[81]     The form required applicants to submit information including extensive business and financial description; a detailed business plan for the proposed registry, including revenue model, market projections, marketing plan, hiring plans, and more; detailed descriptions of technical capabilities, including such matters as billing and collection systems and the times of day in which web-based and telephone support would be available. See Registry Operator's Proposal (Aug. 15, 2000), <http://www.icann.org/tlds/tld-app-registry-operator-proposal-15aug00.htm>.  It required descriptions of the proposed TLD string; the structure of the name space within the proposed TLD;  policies for selection of, and competition among, registrars; plans to protect the interests of trademark owners; policies on data escrow, privacy and Whois; their billing and collection plans; and proposed price and service lists.  TLD Application: Description of TLD Policies (Aug. 15, 2000), <http://www.icann.org/tlds/tld-app-policy-description-15aug00.htm>.  Applicants had to describe the special procedures they would follow to address the expected rush for registration at the TLD's opening, and the trademark-owner protections they would apply then.  Applicants proposing to limit who could register within the domain, or the uses that were to be made of names within the domain, had to describe those criteria, together with the associated application, enforcement, appeal, and cancellation processes.  Id.

[82]     The application form was made available on August 15, and the final application deadline was October 2.

[83]     Id.

(including technical, financial, legal, etc.)."[84]  Staff emphasized that each applicant

> must submit a detailed, multi-part proposal accompanied by extensive supporting documentation.  The effort and cost of preparing a sufficient proposal should not be underestimated. . . . Those who are planning to apply are strongly urged to secure now the professional assistance of technical experts, financial and management consultants, and lawyers to assist in formulation of their proposals and preparation of their applications.

Indeed, staff continued, "your own cost of formulating a proposal and preparing an adequate application will likely be much more" than the $50,000 application fee.[85]  Together with the application form, ICANN released a document describing nine broad values staff would look to in assessing the proposals.[86]

Forty-seven firms filed applications; of those, ICANN returned two for failure to include the $50,000 fee.[87]  Staff's evaluation of the remaining applications was compressed.  The ICANN

---

[84]     TLD Application Process FAQs, FAQ #15, <http://www.icann.org/tlds/tld-faqs.htm>.

[85]     New TLD Application Instructions (Aug 15, 000), sec. I2, <http://www.icann.org/tlds/new-tld-application-instructions-15aug00.htm>.

[86]     Criteria for Assessing TLD Proposals (Aug. 15, 2000), <http://www.icann.org/tlds/tld-criteria-15aug00.htm>.  Successful applications, the document explained, should "preserve the stability of the Internet": They should eliminate or minimize the effects of technical failures in registry or registrar operations, and they should steer clear of anything that challenged ICANN's position as proprietor of the root zone.  Staff would favor TLDs that would help advance the "proof of concept" ICANN sought, providing useful information regarding the feasibility and utility of different types of new TLDs, procedures for launching them, registry-registrar models, business models, and internal policy structures.  See id., points 1 & 2.
    Staff would favor TLDs that would promote competitiveness in the market for registration services; they would favor TLDs that would "sensibly add to the existing DNS hierarchy" and would not confuse users seeking to use the domain name space as a resource-locating tool.  See id., points 3 & 4.  They would favor TLDs that met unmet needs, and that enhanced the diversity of the DNS.  In particular, staff announced its desire to grant applications for "sponsored" TLDs, in which a registry delegated policy-making to "a sponsoring organization that allows participation of the affected segments of the relevant communities."  See id., points 5-7; New TLD Application Process Overview (Aug. 3, 2000), sec. 1(b), <http://www.icann.org/tlds/application-process-03aug00.htm#1b>.  Each application was to incorporate protection ICANN deemed sufficient for trademark holders, and each was required to "demonstrate realistic business, financial, technical, and operational plans and sound analysis of market needs."  See Criteria for Assessing TLD Proposals (Aug. 15, 2000), <http://www.icann.org/tlds/tld-criteria-15aug00.htm>, points 8 & 9.

[87]     See TLD Application Review Update (13 October 2000), <http://www.icann.org/tlds/tld-review-update-13oct00.htm>.  One of these was a proposal for .number, .tel and .phone TLDs from De Breed Holdin B.V.; the other was a proposal for a .wap TLD (for Wireless Application Protocol) from a firm, called the dotWAP Domain Registry, formed for the purpose.

meeting at which the selections were to be made would begin in just six weeks, on November 13. The opportunity for public comment was even more compressed: Members of the public could not comment until the application materials were made available on the Web for the public to see, and that process was significantly delayed. Staff announced that they had posted "most of" the materials by October 23;[88] they reported on November 1 that they had posted all of the "basic" materials, with "a few partial omissions."[89]

On November 10, just one working day before the four-day ICANN meeting was to begin, staff made available its crucial "Report on New TLD Applications."[90] The document incorporated contributions from three outside technical advisors, together with advice from the Arthur Andersen accounting firm and Jones, Day Reavis & Pogue (ICANN's outside counsel).[91] It included a brief summary of each application, consisting of a thirteen-item template for each and a brief summary of any public comments received. This was the first moment that any applicant learned of the staff's assessment of its proposal; staff had declined to meet with applicant representatives at any point during the process.[92]

The body of the report divided the applications into eight categories. Within each category, the report first identified applications that "did not merit further review" because they were deemed unsound for technical or business reasons.[93]  That disposed of sixteen applications. The report discussed the remaining applications in more detail, attempting to compare applications in each category.  In the final analysis, the report described fifteen applications as plausible candidates for going forward; it cautioned, though, that the Board "could responsibly select" only a limited number of them.[94]

---

[88]     Staff attributed the delay to technical issues and disputes with certain applicants over the confidentiality of parts of their applications. See TLD Application Review Update (23 October 2000), <http://www.icann.org/tlds/tld-review-update-23oct00.htm>.

[89]     TLD Application Review Update (1 November 2000), <http://www.icann.org/tlds/tld-review-update-01nov00.htm>.

[90]     <http://www.icann.org/tlds/report/>.  The report carries a November 9 date, but it was not released until November 10.  See Reconsideration Request 00-8, Recommendation of the Committee (Mar. 5, 2001), <http://www.icann.org/committees/reconsideration/rc00-8.htm> ("the Committee is sensitive to the concern that the evaluation team's report was not available to the public – and thus to [the applicant] – until November 10"); Reconsideration Request 00-12, Recommendation of the Committee (Mar. 5, 2001), <http://www.icann.org/committees/reconsideration/rc00-12.htm>; Announcements, <http://www.icann.org/announcements/> (using Nov. 10 date); New TLD Program Application Process Archive, <http://www.icann.org/tlds/app-index.htm> (same).

[91]     See Appendix A: Outside Advisors (Nov. 9, 2000), <http://www.icann.org/tlds/report/report-appa-09nov00.htm>.

[92]     See TLD Application Review Update (23 October 2000), <http://www.icann.org/tlds/tld-review-update-23oct00.htm>.

The staff report kicked off frenzied activity on the part of many of the applicants, as they attempted in the meager time remaining to generate and file comments refuting staff's characterizations of their applications.  On November 15, in a spectacle reminiscent of nothing so much as television's "The Gong Show,"[95] each of the forty-four applicants was given exactly three minutes to appear before the Board, respond to questions, and make its case.  The Board, after all, had allocated only an afternoon to hear the applicants and take public comment; even giving each applicant three minutes (plus enough time to walk to and from the microphone) ate up nearly two hours of that time.  Most of the applicants played along gamely, trying to make the best of their three minutes.  When one applicant used its time to criticize ICANN's "highly flawed process," departing chair Esther Dyson was tart: "I'm really sorry," she said, "we gave you the chance to speak and you did not take very good advantage of it."

Four of the Board members had recused themselves (although they remained on the dais),[96] and three others chose not to participate.[97] The following day, when the Board met to make its decisions, discussion among the twelve members remaining was lively. While ICANN critics had on other occasions worried that the Board's open meetings simply ratified decisions already reached elsewhere, it seemed plain in this case that the Board members had not discussed the applications with each other before.[98] They had a single day's session to make their decisions (along with conducting other, unrelated business), and they were making those decisions from scratch.

---

[93]     Specifically, the report indicated, these applications did not "demonstrate specific and well-thought-out plans, backed by ample, firmly committed resources, to operate in a manner that preserves the Internet's continuing stability," and had not demonstrated realistic business, financial, technical, and operational plans and market analysis.  Somewhat oddly, though, the report stated that this judgment was "comparative" – and that its conclusion that an applicant's technical or business plans were not realistic was therefore "not necessarily a judgment that either the applicant or its proposal had no merit."  Report on New TLD Applications, sec. III.B.1.a, <http://www.icann.org/tlds/report/report-iiib1a-09nov00.htm>.

[94]     The report's favored applications were Afilias (.info, .site, .web), iDomains (.biz, .ebiz, .ecom), JVTeam (.biz), KDD Internet Solutions (.biz, .home), Neustar (.dot, .info, .site, .spot, .surf, .web), CORE (.nom), JVTeam (.per), Sarnoff (.i), Global Name Registry (.name, .nom, .san, .xing), Cooperative League of the USA (.co-op, .coop), International Confederation of Free Trade Unions (.union), Museum Domain Management Association (.museum), Société Internationale de Télécommunications Aéronautiques (.air), World Health Organization (.health), and SRI International (.geo).

[95]     One observer referred to it, more colorfully than I could have, as "the Gong Show spectacle of dozens of sweaty suits having 90 seconds to justify their sorry existence."  NTK now (Nov. 17, 2000), <http://www.ntk.net/?back=2000/now1117.txt>.

The Board's discussion was halting at the outset; the board members had varying views on what they should be doing and how.[99] They settled on an approach in which they would consider the applications one by one, putting the plausible ones into a metaphorical "basket," and returning to the basket when the list was done. Their procedure was anything but well-organized, though; after their initial identification of plausible applications, the Board went back through the applications in their basket multiple times, changing their minds as they went.[100] One director maintained a "parallel basket," containing applications that had not succeeded on the first pass, but which stayed in the running nonetheless.

---

[96] Three of the four – Rob Blokzijl, Greg Crew, and Phil Davidson – had announced their recusal just two weeks before, on November 1. TLD Application Review Update (1 November 2000), <http://www.icann.org/tlds/tld-review-update-01nov00.htm>. Amaddeu Abril i Abril had recused himself on October 2.

[97] All three of the nonparticipating directors (Geraldine Capdeboscq, George Conrades, and Eugenio Triana), along with the recused Greg Crew, were departing; their terms of office expired that day. The only departing director who did participate was outgoing chair Esther Dyson.

[98] "These are the benefits of open process: any ICANNspiracy theories evaporated in the face of the truth, which was as arbitrary and bizarre as anyone could have hoped." NTK now (Nov. 17, 2000), <http://www.ntk.net/?back=2000/now1117.txt>.

[99] For example, many applications proposed more than one string. A Board member asked whether the Board should approve a proposal including all proposed strings, or approve a string and then seek out a suitable proposal. See Scribe's Notes: ICANN Board Meeting - November 16, 2000, sec. VI.C.6 (Sang-Hyon Kyong). Different directors expressed different preferences.

[100] See Ted Byfield, Ushering in Banality, Telepolis (Nov. 27, 2000), <http://www.heise.de/tp/english/html/result.xhtml?url=/tp/english/inhalt/te/4347/1.html&words=ICANN>. Byfield notes "the higgledy-piggledy path traced by the board as it tried to decide what exactly it was deciding on. Was it a specific TLD, a seemingly viable proposal, or a strong applicant? No clear consensus, with inconsistent results." Because the directors had no "coordinated plan or procedure" for picking winners, he charges, "their discussions lurched and reeled from sophomoric ramblings to vacuous platitudes to petty preferences and back again with disorienting rapidity." Id.

Oddnesses seemed to abound. A commercial aviation trade association had applied for the .air TLD, proposing to mirror its content under .aer and .aero. One director questioned whether ICANN could really allocate ".air"; the air, after all, was a public resource. The Board gave the applicant .aero instead.

Another application, from Sarnoff, proposed the .iii TLD string for a personal domain name space (that is, the TLD would issue domain names such as jonweinberg.professor.iii). Well after the Sarnoff application was placed in the basket, and reconfirmed on a second pass, Mike Roberts, ICANN's CEO, objected that the string was unacceptable because it was "unpronounceable" and without semantic meaning. (While one of ICANN's announced selection criteria had suggested a preference for strings with semantic meaning across a wide range of languages, none had indicated that the *sound* of the label when pronounced should be a factor.) Roberts urged that the application be deleted. After discussion, there seemed to be a Board consensus in favor of granting the application either in its original form, or contingent on staff's negotiating with the applicant over an alternate string.[101]

Joe Sims, ICANN's outside counsel, then suggested that the application be denied because Sarnoff had at the last minute agreed to enter into a joint venture with another strong applicant; this created "uncertainties" that cut against granting the application. Louis Touton, ICANN's general counsel, suggested that negotiating with Sarnoff over a new string could be seen by other applicants as unfair. The Board took three additional votes on the application in quick succession; ultimately it was not selected. Watching the process unfold, it was hard to avoid the conclusion that a solid proposal had faded, notwithstanding strong Board support, as a result of concerted opposition from staff.

To a great extent, the Board was handicapped by its self-imposed obligation to make all decisions in a single afternoon and on the fly, without further research or consultation. Faced with the question whether the proposed sponsor of a .travel TLD fully represented the travel industry, Board Chair Esther Dyson urged that the possibility that the sponsor was unrepresentative, whether it was so or not, was enough to doom the application. She explained, according to the scribe's notes, "We're not here to do everything that might make sense if we fully investigate it; we're choosing proof-of-concept domains that don't have these problems."[102]

---

[101]     Sarnoff's application had noted that other possible strings, including .one, would be acceptable.

Perhaps the most confused moments came in connection with the decision what character string to award in connection with the successful application from Afilias. Afilias wanted .web, but that string had long been used by another applicant, which operated a registry accessible via an alternate root.[103] Vint Cerf, ICANN's incoming chair, was sympathetic to that other .web application; finding insufficient support for granting the other application, he urged that .web should instead be "reserved," and that Afilias should receive another string. Cerf then sparred with Touton and Sims over the questions to be voted on, Touton and Sims seeming to formulate those questions so as to favor giving Afilias .web, with Cerf doing the opposite. Several (confusing) votes followed, and Cerf prevailed; Afilias was assigned the .info TLD.

When the day was through, the ICANN Board had approved the opening of negotiations with seven prospective TLD registries. It had not covered itself in glory; the new TLDs were a lackluster lot. It was hard to characterize the afternoon's decision-making process as anything but arbitrary.[104]

Eleven of the disappointed applicants filed petitions for reconsideration. Petitioners urged, among other things, that the staff report contained gross inaccuracies;[105] that ICANN had given applicants no meaningful opportunity to respond to the report or to make their cases;[106] that the selection criteria were vague and subjective;[107] that the Board sandbagged applicants by rejecting applications on the basis of unannounced criteria;[108] and that the Board's consideration was arbitrary, treating similarly situated applicants differently.[109]

---

[102]    See Scribe's Notes: ICANN Board Meeting - November 16, 2000, sec. X.F.4.

[103]    On alternate roots, see supra notes 11-12 and accompanying text.

[104]    As Christopher Chiu of the ACLU put it: "They tried to become an arbiter, and all they did was make arbitrary decisions." Quoted in Aaron Pressman, ICANN: 7 Out of 44 Ain't Bad, The Industry Standard (Nov. 16, 2000), <http://www.thestandard.com/article/0,1902,20272,00.html>.

[105]    See, e.g., Reconsideration Request of Abacus America (Nov. 28, 2000) <http://www.icann.org/committees/reconsideration/vachovsky-request-28nov00.htm>; Reconsideration Request of International Air Transportation Association (Dec. 15, 2000), <http://www.icann.org/committees/reconsideration/goldberg-request-15dec00.htm>; Reconsideration Request of .TV Corporation (Dec. 15, 2000), <http://www.icann.org/committees/reconsideration/dottv-request-15dec00.htm>; Reconsideration request of Image Online Design, Inc. (Dec. 15, 2000), <http://www.icann.org/committees/reconsideration/iodesign-request-15dec00.htm>; Reconsideration Request of ICM Registry (Dec. 16, 2000), <http://www.icann.org/committees/reconsideration/icm-request-16dec00.htm>; Reconsideration Request of Monsoon Assets Limited (May 2, 2001),

ICANN rejected all of the petitions, issuing a remarkable statement that in important extent conceded the failings that petitioners complained of. That the selection criteria and the ultimate judgments were subjective, ICANN explained, was not a flaw to task it with; that subjectivity was "inherent" in the process.[110] It was "clearly articulated from the beginning of the process" that similar proposals could be treated differently.[111] Moreover, it was not a sufficient basis for reconsideration that "there were factual errors made, or there was confusion about

---

<http://www.icann.org/committees/reconsideration/knight-request-2may01.htm>.

[106]    See, e.g., Reconsideration Request of Abacus America, supra n. 105; Reconsideration Request of .TV Corporation, supra n. 105; Reconsideration Request of .Kids Domains, Inc. (Feb. 1, 2001), <http://www.icann.org/committees/reconsideration/howe-request-1feb01.htm>.

[107]    See, e.g., Reconsideration Request of .TV Corporation, supra n. 105.

[108]    See, e.g., Reconsideration Request of International Air Transportation Association, supra n. 105; Reconsideration Request of Sarnoff Corp. (Dec. 15, 2000), <http://www.icann.org/committees/reconsideration/sarnoff-request-15dec00.htm>; Reconsideration Request of .Kids Domains, Inc., supra n. 106.

[109]    See, e.g., Reconsideration Request of International Air Transportation Association, supra n. 105; Reconsideration Request of Image Online Design, Inc., supra n. 105.

[110]    Reconsideration Request 00-8 (Abacus America): Recommendation of the Committee (Mar. 5, 2001), <http://www.icann.org/committees/reconsideration/rc00-8.htm>; see also Reconsideration Request 00-12 (.TV Corporation): Recommendation of the Committee (Mar. 16, 2001), <http://www.icann.org/committees/reconsideration/rc00-12.htm>. The reconsideration committee incorporated its statement in *Reconsideration Request 00-8 (Abacus America)*, by reference, into all of the reconsideration decisions it rendered in connection with the new TLD process.

[111]    Reconsideration Request 00-13 (Image Online Design): Recommendation of the Committee (Mar. 16, 2001), <http://www.icann.org/committees/reconsideration/rc00-13.htm>. This seems extravagant. ICANN staff did make clear at the outset that they intended the process to generate only a small number of TLDs, so that worthwhile TLD applications might not be granted. They did not, however, state that they anticipated the process to be arbitrary. The statement in ICANN's New TLD Application Instructions that "[o]nly a limited number of TLDs will be established in this round of applications, and it is likely that only applications with very high qualifications will be accepted" better exemplifies staff's initial description of the process. <http://www.icann.org/tlds/new-tld-application-instructions-15aug00.htm>.

various elements of a proposal, or each member of the Board did not fully understand all the details of some of the proposals."[112]  After all, given the subjective and fact-intensive nature of the evaluation, *any* process – even one unmarred by confusion and error – would yield results on which reasonable people could differ.  That reasonable people could conclude that other selections would have better advanced ICANN's goals was simply "inevitable."[113]

Moreover, ICANN continued, the Board could not be faulted for departing from the announced selection criteria.  Those criteria were never "intended to be a rigid formula for assessing the merits of TLD proposals" -- they were simply drafting guides for the applicants.[114]  Finally,  because ICANN's goal was proof of concept, it had never intended to treat "the absolute or relative merit of any application [as] the single factor determining the outcome."[115]  To the extent that the Board had passed over more meritorious applications in favor of less meritorious ones, that was simply irrelevant.[116]

---

[112]     Reconsideration Request 00-8 (Abacus America): Recommendation of the Committee, supra n. 110.

[113]     Id.

[114]     Reconsideration Request 01-2 (.Kids Domains): Recommendation of the Committee (Apr. 30, 2001), <http://www.icann.org/committees/reconsideration/rc01-2.htm>.

[115]      Reconsideration Request 00-14 (SRI International): Recommendation of the Committee (Mar. 16, 2001), <http://www.icann.org/committees/reconsideration/rc00-14.htm>.

[116]     Turning to more concrete process concerns, ICANN stated that the weekend and three working days between the staff report's release and the Board's decision provided applicants with sufficient opportunity to respond to any errors – and all applicants, after all, were subject to the same hurried schedule. Reconsideration Request 00-8 (Abacus America): Recommendation of the Committee, supra n. 110; Reconsideration Request 00-10 (Group One Registry): Recommendation of the Committee (Mar. 5, 2001), <http://www.icann.org/committees/reconsideration/rc00-10.htm>; Reconsideration Request 00-12 (.TV Corporation): Recommendation of the Committee, supra n. 110.  (Though ICANN repeatedly states that the process was fair, this answer does suggest that ICANN deemed it less important that the process be accurate and reliable, than that it be equally inaccurate and unreliable for all.)  The three minute dog-and-pony-show, ICANN stated, was appropriate because "[t]he opportunity to make a presentation at the public forum was simply the final step in an extensive process, available so that any last-minute questions could be asked or points made."  Reconsideration Request 00-9 (International Air Transportation Association): Recommendation of the Committee (Apr. 30, 2001), <http://www.icann.org/committees/reconsideration/rc00-9.htm>.  Indeed, it "re-emphasized [ICANN's] commitment to maximum transparency," by making clear (if only after the fact) that all input to the process from applicants needed to have been in writing, "so that the entire Internet community would have the opportunity to read it, consider it, and respond to it."  Reconsideration Request 00-12 (.TV Corporation): Recommendation of the Committee, supra.

It is hard to know what to do with such an extraordinary explanation. ICANN tells us here that the selection process it chose was so inherently subjective, so much in the nature of a crap-shoot, that there is simply no point in identifying errors in its consideration of the applications; any such errors are simply irrelevant. This statement is not accompanied by any abashedness, by any suggestion that such a process might be inherently flawed. The ad hoc, subjective nature of the process, rather, is presented as a feature and not a bug. It is presented as the only possible path available to ICANN to initiate the proof of concept. Indeed, a later ICANN statement suggested, ICANN's public trust *demands* that it add TLDs to the root only through processes like these, in which the Board, with the applications before it, endeavors to make those selections that best achieve the larger public interest as the Board perceives it.[117]

### III. CONCLUSION

A number of things are notable about the history I have just told. The story begins with Jon Postel's proposal to expand the name space. That proposal contemplated hundreds of new top level domains, and an administrative process that was lightweight in two respects. First, applicants for new TLDs could get them without jumping through complicated procedural hoops. Second, the applicants would not have to satisfy onerous substantive standards. Precisely because Postel proposed to make many new TLDs available, he did not need to limit the universe of those who applied.

Postel's proposal ran into immediate opposition from business groups. They feared the consequences of quick domain name expansion for trademark owners. More broadly, players sympathetic to business concerns raised questions under the banner of "Internet stability" – if many new registries were easily formed, might not some fail?  Would that dampen the consumer's enthusiasm for e-commerce?  Might not some consumers be confused by the multiplicity of new domains, again making the Internet less hospitable to buying and selling?

---

[117]     In ICP-3: A Unique, Authoritative Root for the DNS (July 9, 2001), <http://www.icann.org/icp/icp-3.htm>, ICANN urges that it would be inappropriate to include any gTLD in the root where the particular gTLD has not been subjected to "tests of community support and conformance with consensus processes – coordinated by ICANN." The policy statement states that ICANN would betray the public trust were it, in introducing new TLDs, to place positive value on the fact that a particular applicant was already operating in an alternate root, for that would derogate ICANN's own selection process. ICANN may introduce a particular new gTLD only once the gTLD has been confirmed through "the community's processes," and only where doing so serves the public interest.

At the same time (and largely in response to the fears stirred up by Postel's proposal and the events that followed), the United States government was restructuring the mechanisms of Internet governance.  ICANN was striking, in comparison with IANA,  in the increased representation it gave business interests.  IANA was controlled by the technical elite; one of its functions was to serve as editor for a key series of documents generated by the Internet Engineering Task Force.  By contrast, ICANN empowered business users: Its Names Council was nothing but representatives of various industry groupings.  Operating in a world in which business and governments had woken up to the importance of the domain name space, and "working within the system to balance competing interests, many of which possess economic power,"[118] ICANN showed great sensitivity to business concerns.

In putting forward proposals to expand the name space, therefore, ICANN's approach was far different from Postel's.  It emphasized that only a few lucky applicants would be allowed in, and only as a "proof of concept."  It imposed extensive threshold requirements for even considering the application, in an attempt to ensure (in part) that no new TLD registry would fail or suffer difficulties, thus threatening "Internet stability."  And it selected the lucky winners through a process designed to give it the greatest degree of control over the ultimate outcome, notwithstanding the dangers of subjectivity and arbitrariness inherent in that approach.

This paper is part of an ongoing examination of ICANN and its relationship with the top level domain registries.  In a later paper, I will examine the striking parallels between ICANN's comparative process in this case and the FCC's now-abandoned comparative hearing process for broadcast licenses.  Both processes are usefully examined as examples of ad hoc, situationally-sensitive rather than rule-based decision making.  A variety of other issues ICANN has confronted in this process are similarly familiar to lawyers familiar with FCC processes.

---

[118]    Michael Roberts (then ICANN's CEO), Comments on the Civil Society Statement (July 30, 2000), <http://www.cpsr.org/internetdemocracy/Statement_July-13_comments.html>.

ICANN's selection of seven registries, further, was not the end of the story. I will examine the command-and-control regulation ICANN has imposed through its negotiation of contracts with the new registries it has chosen. The new contracts give ICANN closely detailed control over the new registries and their business models. Here, too, I will draw parallels – and note contrasts – with the FCC's experience. In many ways, I will argue, ICANN is picking the worst of the FCC history to adopt. Critics has ruthlessly criticized the FCC processes, most of which that agency has now abandoned; ICANN has effortlessly managed to surpass the worst that the FCC ever approached.

Most of ICANN's regulatory imperative derives from its decision to maintain scarcity of top level domains, rolling them out only slowly. (Once again, the parallel with the FCC is instructive.) I will assess this decision. To what extent were – and are -- alternatives feasible?